# *In vitro* degradation and cytotoxicity response of biobased nanoparticles prepared by thiol-ene polymerization in miniemulsion

Fernanda Hoelscher [a], Priscila B. Cardoso [a], Graziâni Candiotto [b], Camila Guindani [a, c], Paulo Feuser [a], Pedro H. H. Araújo [a] and Claudia Sayer [a]

[a] Department of Chemical Engineering and Food Engineering - Federal University of Santa Catarina - EQA/UFSC - C.P. 476, CEP 88040-900, Florianópolis, SC, Brazil.

[b] Institute of Chemistry – Federal University of Rio de Janeiro – IQ/UFRJ, CEP 21941-909, Rio de Janeiro, RJ, Brazil.

[c] Chemical Engineering Program, COPPE, Federal University of Rio de Janeiro - PEQ/COPPE/UFRJ, CEP 21941-972, Rio de Janeiro, RJ, Brazil.

*Corresponding author: claudia.sayer@ufsc.br

**ABSTRACT**

Biodegradability is a key feature for the application of polymeric devices in medicine. This study reports an experimental and theoretical study of the degradation of poly(thioether-ester) (PTEe) nanoparticles in aqueous media. The α,ω-diene diester derived from vegetable oil, 1,3-propylene diundeca-10-polenoate (Pd10e) was used as monomer in the solvent-free synthesis of Pd10e-based nanoparticles (A-PTEe nanoparticles) via thiol-ene miniemulsion polymerization. The theoretical partition coefficients of A-PTEe and a PTEe based on dianhydro-D-glucityl diundec-10-enoate (DGU) (B-PTEe nanoparticles) were calculated using density functional theory (DFT), in order to compare their degradation behavior. The results showed that A-PTEe is more hydrophilic than B-PTEe, thus indicating the possible faster degradation of the former. The experimental degradation studies showed that, in fact, A-PTEe nanoparticles are faster degraded than B-PTEe, presenting substantial molecular weight decrease, which confirms the theoretical results. The effects of degradation could be observed in the chemical composition and thermal properties of the polymer. Considering its applicability potential as a biomaterial due to its fast degradation behavior, the cytotoxicity of A-PTEe nanoparticles and its degradation products were evaluated. *In vitro* assays confirmed the biocompatibility of A-PTEe nanoparticles and its degradation products when exposed on fibroblasts and red blood cells. These results suggest A-PTEe nanoparticles can be promising candidates as biobased nanocarriers for biomedical applications.

# 1. INTRODUCTION

Biodegradable synthetic polymers have been used in a wide range of applications where its presence is necessary only for developing a function during a certain period of time. These applications include temporary implants for tissue engineering, drug delivery, and applications in the packaging sector, helping in the decrease of plastic waste pollution [1]. In this context, the synthesis of nanoparticles by thiol-ene addition reactions in miniemulsion is still an underexplored area, especially for polymers obtained by polymerization reactions using renewable monomers. These reactions are proving to be promising strategies in the pharmaceutical and biomedical fields, allowing the green synthesis of biocompatible and biodegradable polymers and the introduction of functional groups in the polymeric chain (e.g. glycosidic, ether, ester and amide bonds, besides the incorporation of the sulfide group). Some of these groups, such as ester functionality, can be readily hydrolyzed, being potentially biodegradable [2–6].

Some recent works reported the polymerization via thiol-ene addition reactions using renewable monomers. Machado et al. [7] performed the miniemulsion polymerization from the renewable monomer dianhydro-D-glucityl diundec-10-enoate (DGU) with 1,4-butanedithiol. Poly(thioether-ester) (PTEe) nanoparticles based on DGU, were used to encapsulate clove oil with improved antioxidant activity [8] and magnetic nanoparticles for hyperthermia studies [9].

The degradation behavior is a very important feature of polymer nanoparticles for applications as nanocarriers for sustained drug delivery, which indeed can differ a lot in comparison to the degradation behavior of larger devices or microparticles made of the same polymer [10]. In their work, Hoelscher et al. [11] studied the degradation of PTEe nanoparticles prepared with the monomer DGU (here called B-PTEe). B-PTEe nanoparticles showed interesting biodegradation characteristics, especially when exposed to acidic and enzymatic solutions, presenting molecular weight decrease of 37% (3 months, acidic solution) and 90% (240 h, enzymatic solution).

The aim of the present work was to investigate the degradation behavior of the vegetable oil-derived PTEe nanoparticles prepared by thiol-ene miniemulsion polymerization of the monomer 1,3 propylene-diundeca-10-enoate (Pd10e), in a solvent-free system. These Pd10e based nanoparticles were henceforth called A-PTEe. The degradation was evaluated in enzymatic solution, acidic solution and in a phosphate-buffered solution (PBS), and the material was characterized in terms of functional groups, average intensity particle size (Dp), dispersity (Đ), molecular weight and thermal properties. By the help of density functional theory (DFT), a theoretical study was performed comparing A-PTEe and B-PTEe structures, providing deeper understanding on the stability and degradation behavior of the PTEe-based materials investigated. Fig. 1 shows the polymerization reaction and compares the chemical structure of the poly(thioether-ester)s based on (a) Pd10e monomer (A-PTEe), and (b) DGU monomer (B-PTEe). Since the main application focus of these nanoparticles is on the sustained drug release for medical treatments, cytotoxicity assays of the A-PTEe degraded nanoparticles were performed, using murine fibroblast cells (L929). Hemolysis assays were also performed, in order to evaluate the biocompatibility of the nanoparticles' degradation products with human red blood cells (RBCs).

These are original results on the degradation behavior and cytotoxic characteristics of PTEe nanoparticles, and we consider it an important contribution to the development of new biobased drug delivery systems with enhanced degradability characteristics.

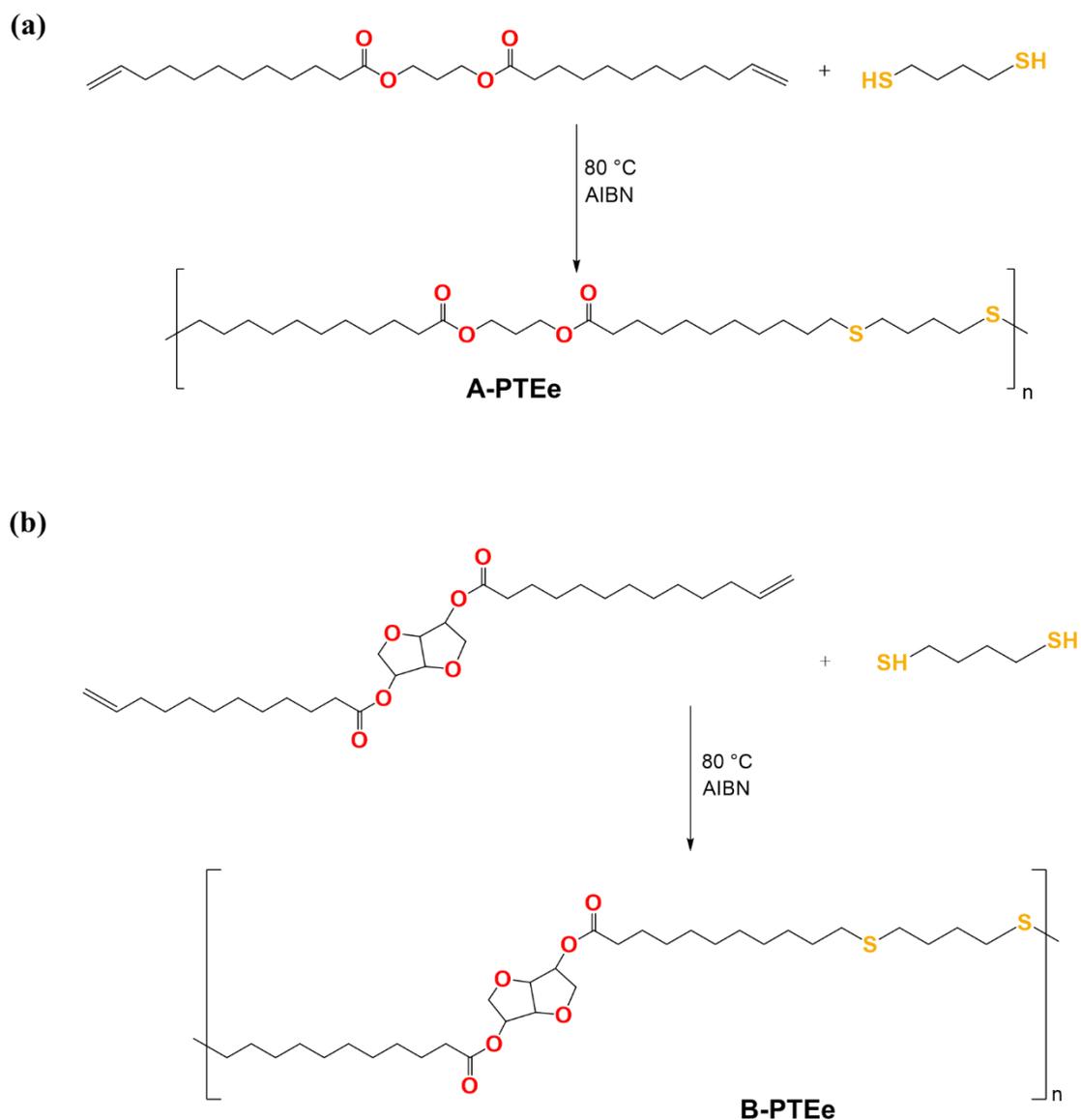

**Fig. 1** Synthesis of the thioether-esters polymers produced using as monomers: (a) α,ω-diene diester 1,3-propilene diundec-10-enoate and 1,4-Butanedithiol, herein called A-PTEe, and (b) a,ω-diene diester dianhydro-D-glucityl diundec-10-enoate and 1,4-Butanedithiol, herein called B-PTEe

## 2. MATERIAL AND METHODS

### 2.1. Material

For synthesis of the A-PTEe nanoparticles the followed reagents were used: Lutensol AT 80 (BASF); 1,4-Butanedithiol (Bu(SH)$_2$); α, ω- diene diester 1,3-propylene diundec-10-enoate, a monomer derived from vegetable oil, synthetized by Cardoso et al. [12]; and azobisisobutyronitrile (AIBN). For degradation assays, the following reagents were used: hydrochloric acid (HCl, P.A 36.5%); sodium hydrogen phosphate (NaHPO$_4$, P.A 98%), sodium dihydrogen phosphate (NaH$_2$PO$_4$, P.A 99%) and the free enzyme Candida Antarctica lipase B (CalB - Novozymes Latin América) previously concentrated. In order to determine the enzyme activity, p-nitrophenyl palmitate (pNPP), sodium carbonate (Na$_2$CO$_3$) were used. The GPC analysis were perfomed with tetrahydrofuran (THF, Vetec). Distilled water was used in all the experiments.

**2.2.    Synthesis of A-PTEe nanoparticles**

A-PTEe nanoparticles synthesis were carried out by miniemulsion polymerization technique, according to a formulation optimized by Cardoso et al [12]. Scheme 1 illustrates the formation of A-PTEe nanoparticles. Typically, 0.6 g of surfactant Lutensol AT80 (MW 3.8 kDa, BASF) were solubilized in 20 g of distilled water for 5 min (150 rpm) to prepare the aqueous phase. To prepare the organic phase, 0.008 g of AIBN initiator were homogenized in 2 g of the monomer Pd10e for 15 min (150 rpm). Afterwards, the organic phase was mixed to the aqueous phase under magnetic agitation (300 rpm) and then 0.6 mL of the Bu(SH)$_2$ was added to the system and homogenized (Pd10e:Bu(SH)$_2$ 1:1 ratio). The emulsion formed was sonicated with the help of a Sonic Dismembrator (Fisher Scientific, model 500) and a 1/2'' tip in an ice bath, for 4 minutes and 60% of amplitude, pulse regime of 10 s with 2 s of pause. After sonication, the emulsion was placed in a thermostatic bath at 80°C and polymerization reaction was carried out for 4 hours. Miniemulsions were stored in refrigerator (4 °C) until characterization and degradation assays were performed.

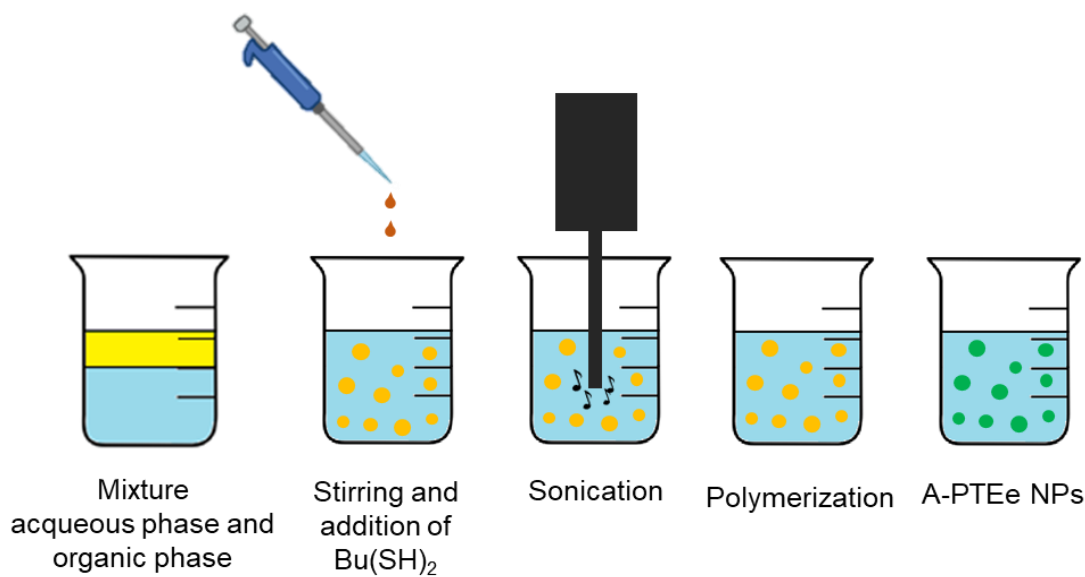

**Scheme 1** Production of A-PTEe NPs by miniemulsion polymerization technique.

## 2.3. Characterization of A-PTEe nanoparticles

*Particle size and dispersity*: The particle size was measured by dynamic light-scattering (DLS). One drop of the sample was diluted in 1.5mL of distilled water in a glass cuvette. The particle size in intensity (Dp) and dispersity (Đ) were measured by a dynamic light scattering (DLS) method using a Zetasizer Nano ZS (Malvern Instruments).

*Attenuated Total Reflectance Fourier Transform Infrared Spectroscopy (ATR-FTIR)*: The miniemulsion samples were dried and analyzed by ATR-FTIR (Bruker, model TENSOR 27) using a ZnSe crystal from PIKE, in a resolution of 4 cm$^{-1}$, wavenumber range of 4000 cm$^{-1}$ – 600 cm$^{-1}$.

*Differencial Scanning Calorimetry (DSC)*: DSC analysis were performed using a Perkin-Elmer equipment (model 400) under inert atmosphere , with a nitrogen flow of 50 mL/min, from -10 to 150 ºC, with a heating rate of 10 ºC /min and $T_m$ was obtained in the second heating run through the endothermic peaks. 10 mg of dry samples were used for each analysis.

*Gel Permeation Chromatography (GPC)*: Molecular weight distribution (MWD), number average molecular weight ($M_n$) and weight average molecular weight ($M_w$) were determined by GPC. The measurements were performed in THF with a LC-20A gel permeation chromatography system (Shimadzu) equipped with PLgel 5µm Mini MIX-C Guard 50 x 4 mm guard column and two columns packet with styrene-divinylbenzene copolymers spheres Mini PLgel MIX 2504, 6mm arranged in series. For the analysis, 0.02 g of polymer was dissolved in 4 mL of THF and filtered with a nylon syringe filter with pore of 0.45 mm and diameter 25 mm. The eluent THF was used at 40 °C with a volumetric flow rate of 0.3 mL/min. Calibration was achieved using polystyrene standards ranging from 580 to 9.8 $10^6$ Da.

*Proton Nuclear Magnetic Resonance ($^1H$ NMR):* $^1$H NMR spectroscopy was performed on a Bruker Avance 300, operating at 300 MHz. All spectra were referenced internally to residual proton signals of the deuterated solvent. Samples were solubilized in CDCl3 ($\delta$ = 7.27 ppm for $^1$H NMR).

## 2.4. Computational methods

The molecular structures of copolymers A-PTEe and B-PTEe used for computational simulation were generated and analyzed by Avogadro [13] version (1.2.0).

First, calculations related to the geometry optimization of A-PTEe and B-PTEe structures were performed, allowing the determination of properties such as the dipole moments, Gibbs free energy ($G$) and partition coefficient ($P$). The molecular structures of the polymers were optimized in the gas-phase, water and n-octanol. The optimized structures were confirmed as real minima by vibration analysis (no imaginary frequency was detected). The results were obtained at normal temperature and pressure (NTP) using B3LYP hybrid functional of DFT in conjunction with 6-311++G** basis set. The Gibbs free energy of solvation in water and n-octanol for the polymers were computed using solvation model based on electronic density (SMD). All DFT calculations were performed using the gaussian 09 package program.

**2.5.     *In vitro* degradation of A-PTEe nanoparticles**

The degradation assays were performed by applying acidic, enzymatic and buffer solutions, at 37 °C, in a bacterial oven following the procedure described by Hoelscher et al. [11]. In this assays, 1 mL of the miniemulsion containing A-PTEe nanoparticles was added to 2 mL of hydrochloric acid (pH 2.8, 0.01 M) or PBS solution (pH 7.4, 0.2 M). In the case of enzymatic degradation, the samples were added to a PBS solution containing CalB 7% (w/w) enzyme. The enzyme concentration was performed according to Chiaradia et al. [14]. The enzymatic activity was performed following Chiou and Wu [15]. For the enzymatic solution, the degradation was carried out until the complete degradation of the nanoparticles and on the other two cases for 3 months. The degraded samples were picked up over predetermined time intervals and physical-chemical analysis, such as molecular weight, functional groups, thermal analysis and particle size were performed. At the end of degradation assays, degradation products were collected and their cytotoxicity was evaluated.

**2.6.     *In vitro* cytotoxicity studies**

2.6.1.   Cytotoxicity assay of the A-PTEe nanoparticles and their degradation products

Murine fibroblast (L929) cell line was selected to evaluate the cytotoxicity of PTEe nanoparticles. Cells were cultured in Dulbecco's Modified Eagle Medium (DMEM) supplemented with 10 wt.% of fetal bovine serum (FBS) and 100 U/mL penicillin-streptomycin. L929 cells were maintained at 37 °C in a humidified incubator containing 5 wt.% $CO_2$. After incubation, cells were seeded at $10^4$ cells per well in 96-well plate.

After 24 h, cells were treated with a medium containing PTEe nanoparticles (resuspended in PBS), at four different concentrations: 50, 100, 150, and 200 µg/mL. After 3 h of incubation the cells were washed three times with PBS and incubated for 24h with fresh medium. After the incubation period, the cell culture medium was removed and the 3-(4,5-dimethylthiazol-2-yl)-2,5-diphenyltetrazolium bromide (MTT) cell viability assay was performed. Briefly, 100 µL of MTT was added and the cells were incubated for 3 h (5% $CO_2$, 37 °C). Afterwards, 100 µL of DMSO was added to dissolve the formazan crystals and the absorbance was measured at 570 nm using a SpectraMax M3, microplate reader. Experiments were performed in triplicate with three wells for each condition and the results were expressed as the percentage of viable cells in comparison to the control group (only cells).

2.6.2. Hemolysis assay

Hemolysis assay was performed following the methodology described by Feuser et al. [16]. The human red blood cells (RBCs) obtained from healthy donors were collected in tubes containing 3.2 wt.% of sodium citrate from three volunteers. In sequence, RBCs were isolated from serum by centrifugation at $1500 \times g$ for 5 min. The RBCs were further washed three times with saline solution and diluted in 2 mL of saline solution. After, 60 µL of the diluted RBCs suspension was added to 940 µL of saline solution. The RBCs were treated with A-PTEe nanoparticles at concentrations of 50, 100, 150 and 200 µg/mL by gentle stirring at 37 °C for 120 min (samples prepared in triplicate). Subsequently, the RBCs suspension was centrifuged at $10.000 \times g$ for 5 min and 100 µL of supernatant from the sample tube were transferred to a 96-well plate. The absorbance value was measured at 540 nm. As positive and negative controls, 60 µL of the diluted human red blood cells suspension was incubated with 940 µL of distillated water and saline, respectively.

# 3. RESULTS AND DISCUSSION

## 3.1. Theoretical degradation study

Computational simulations were performed applying DFT calculations in order to estimate important properties of A-PTEe and B-PTEe, such as Gibbs free energy of solvation and partition coefficient. The theoretical estimation of these properties helps on the prediction of the degradation kinetics of the polymers. The molecular structures of A-PTEe and B-PTEe were optimized using DFT (see Fig. S1, support information). In order to simplify the simulation process, the structures of both polymers were considered as dimmers. In this case, A-PTEe dimmer consisted of one unit of Pd10e and one unit of Bu(SH)$_2$ (Pd10e-Bu(SH)$_2$) and B-PTEe dimmer consisted of one DGU unit and one unit of Bu(SH)$_2$ (DGU-Bu(SH)$_2$). Guindani et al. [17] showed in their study that the use of this kind of simplification provides satisfactory approximation and allows the comparison of the degradation behavior of different materials.

3.1.1. Theoretical octanol-water partition coefficient of A-PTEe and B-PTEe

The DFT simulations allow to determine parameters such as Gibbs free energy ($G$) for the molecular systems in different solvents. The Gibbs free energy of solvation ($\Delta G_{solv}$) is a parameter used to determine the partition coefficient (P), that provides information about hydrophilicity of the system therefore it is a good parameter to understand the degradation behavior of A-PTEe and B-PTEe. The partition coefficient is defined as being the ratio between concentrations of a solute in two phases of a mixture of two immiscible solvents at equilibrium [18] and is often represented in the logarithm form ($log\ P$). To determine the Gibbs free energy of solvation for the polymers, their molecular structures were optimized in the gas-phase, and also in water and n-octanol. The theoretical prediction of partition coefficient logarithm for n-octanol/water mixture ($log\ P^{O/W}$) was calculated according to Equation (1) [17]:

$$log\ P^{O/W} = \frac{\Delta G_{solv}^{W} - \Delta G_{solv}^{O}}{2.303RT} \quad (1)$$

In this equation $\Delta G_{solv}$ is the Gibbs free energy of solvation for two different phases, where the superscript labels $W$ and $O$ are respectively the water and n-octanol solvents, R is the gas constant (1.987 cal K$^{-1}$ mol$^{-1}$) and T is the room temperature (298.150 K). Table 1 presents the results obtained for Gibbs free energy and Gibbs free energy of solvation in water and in n-octanol. The results obtained show that the A-PTEe dimer has a lower partition coefficient value than that obtained for the B-PTEe dimer, which is an indicative that A-PTEe is more hydrophilic than B-PTEe, and is consequently more susceptible to degradation in aqueous media.

**Table 1** Partition coefficient parameters calculated for the A-PTEe and B-PTEe dimers at NPT using DFT/B3LYP/6-311++G** with water and n-octanol solvents in SMD model

| Solvent | Parameter | A-PTEe | B-PTEe |
| --- | --- | --- | --- |
| Gas-Phase | $G$ (Kcal/mol) | 441.717 | 469.515 |
| Water | $G_{solv}^{W}$ (Kcal/mol) | 448.772 | 470.6552 |
| | $\Delta G_{solv}^{W}$ (Kcal/mol) | 7.054 | 1.141 |
| n-Octanol | $G_{solv}^{O}$ (Kcal/mol) | 441.719 | 469.516 |
| | $\Delta G_{solv}^{O}$ (Kcal/mol) | 0.001883 | 0.001255 |
| n-Octanol/water | $log\ P^{O/W}$ | -5.161 | -0.834 |

## 3.2. Experimental degradation study

A-PTEe nanoparticles were synthesized through thiol-ene miniemulsion reactions with the renewable monomer Pd10e, as previously described by Cardoso et al. [7]. A-PTEe presented average molecular weight values of $M_w$ =38 kDa and $M_n$ = 18.1 kDa, while the average particle diameter and the dispersity of the nanoparticles were respectively $D_p$ = 145 ± 2 nm and Đ = 0.19 ± 0.01, presenting unimodal particle size distribution. TEM images of A-PTEe nanoparticles stabilized with SDS were obtained by Cardoso et al. [12]. Both the particle size and molecular weight obtained was very similar to the values obtained in the present work. A-PTEe nanoparticles were then submitted to a degradation study in three different incubation media: in buffer solution (PBS, pH 7.4, 0.2 M), acidic solution (HCl, pH 2.8, 0.01 M) and an enzymatic solution (CalB 7% w/w, enzyme activity 2 U/g).

Fig. 2 compares the degradation curves of A-PTEe and B-PTEe (obtained previously by Hoelscher et al. [11]) in terms of the reduction in molecular weight over several days, under different degradation media. When exposed to enzymatic solution (Fig. 2a), after 30 min of incubation A-PTEe nanoparticles presented a reduction of 90% on its molecular weight. After this abrupt decrease in the first 30 min, the molecular weight maintained a stable value for the next 4.5 h. In the degradation study of B-PTEe [11] (Fig. 2c), it was necessary at least 120 h for the same reduction of 90% in molecular weight.

When incubated in acidic solution (Fig. 2b), the decrease in the molecular weight happened more gradually, reaching a reduction of 51% in 3 months, but still being degraded faster than B-PTEe, which had a reduction of 37% in 3 months of study [11] (Fig. 2d). During the 3 months of study A-PTEe samples did not present changes in the molecular weight up when exposed to PBS solution (support information, Table S1). B-PTEe was reported to present the same behavior [11].

Degradation in enzymatic solution resulted in higher degradation rates, in comparison to the degradation carried out in PBS and HCl solutions. *Candida antarctica* Lipase B can catalyze the cleavage of ester bonds, resulting in faster degradation and weight loss of the polymer [19]. Acidic solutions are also reported to catalyze the hydrolysis of ester bonds, but in a less efficient way, when compared to the enzymatic solutions [20].

When analyzing together the experimental and theoretical data for the degradation of A-PTEe and B-PTEe, it is possible to conclude that the higher hydrophilicity of A-PTEe is a determinant factor that boost the degradation rates of A-PTEe.

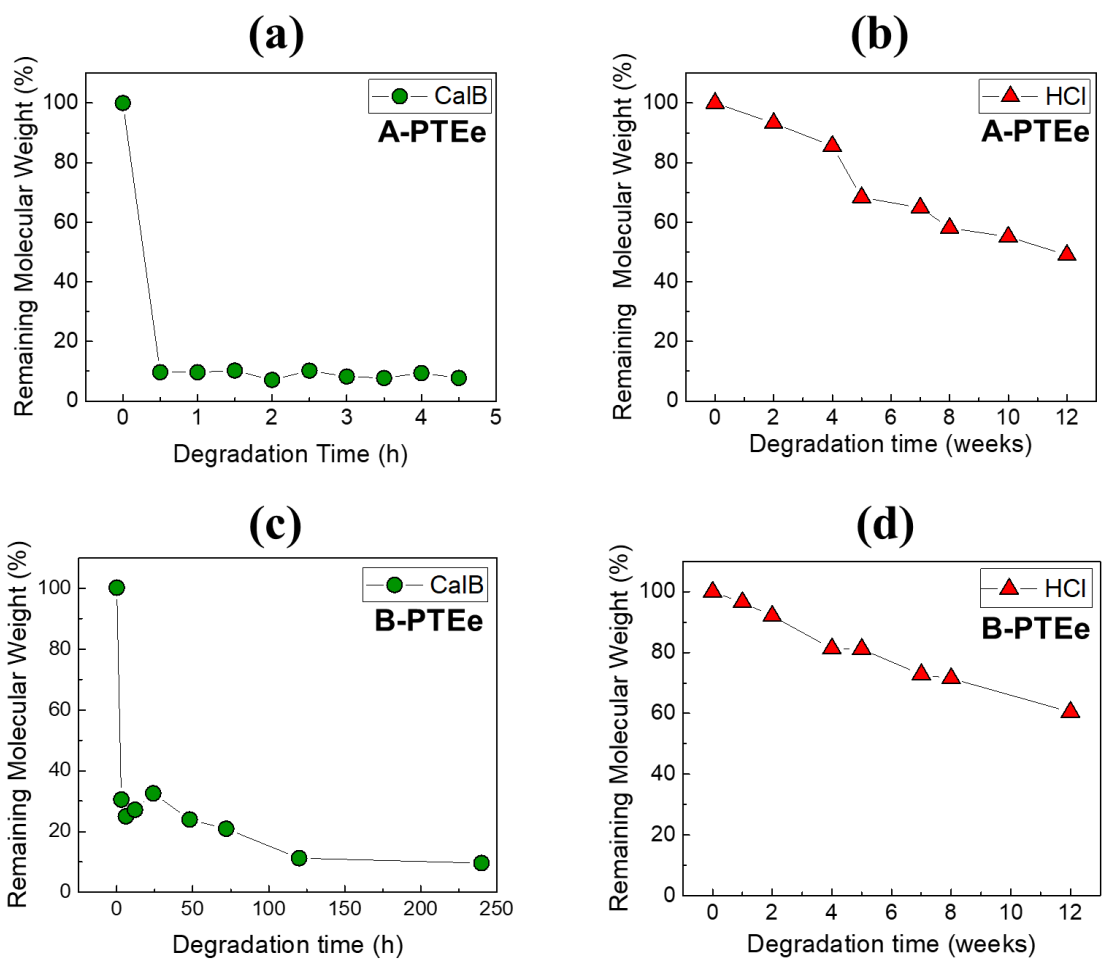

**Fig. 2** Evolution of normalized weight average molar mass ($M_w$) of A-PTEe nanoparticles during degradation in (a) enzymatic solution (CalB) and (b) acidic solution (HCl). Hoelscher et al. [11] performed the same investigation for B-PTEe nanoparticles in (c) enzymatic solution (CalB) and (d) acidic solution (HCl)

Molecular weight distribution of A-PTEe nanoparticles degraded in enzymatic solution is shown in Fig. S2 of support information. The degradation processes have also an effect on the particle size and particle size distribution, as observed in Fig. 3. For the nanoparticles incubated in enzymatic solution (Fig. 3a), it is possible to observe an increase in $D_p$ from $145 \pm 2$ nm to $409 \pm 62$ nm after only 4.5 h of incubation. The same trend is observed for the dispersity (Fig. 3c), which increased from $Đ = 0.06 \pm 0.01$ to $0.8 \pm 0.11$. This increase in particle size upon degradation can be explained by the swelling of the particles. One of the possible causes for this swelling is the formation of acidic degradation products that may favor the entry of water in the polymeric particles, which get also more porous due to erosion caused by degradation [21, 22]. Comparing the degradation behavior in enzymatic medium for A-PTEe and B-PTEe [11], the degradation of A-PTEe was more pronounced. As shown in the DFT simulations herein present, the A-PTEe has strong hydrophilic character when we compare it with the B-PTEe structure derived from DGU monomer, which also proves the behavior shown experimentally.

As observed in Fig. 3b and Fig. 3d, the particles incubated in acidic and PBS solutions remained with constant particle size and dispersity during the entire incubation period, being in accordance with the molecular weight results.

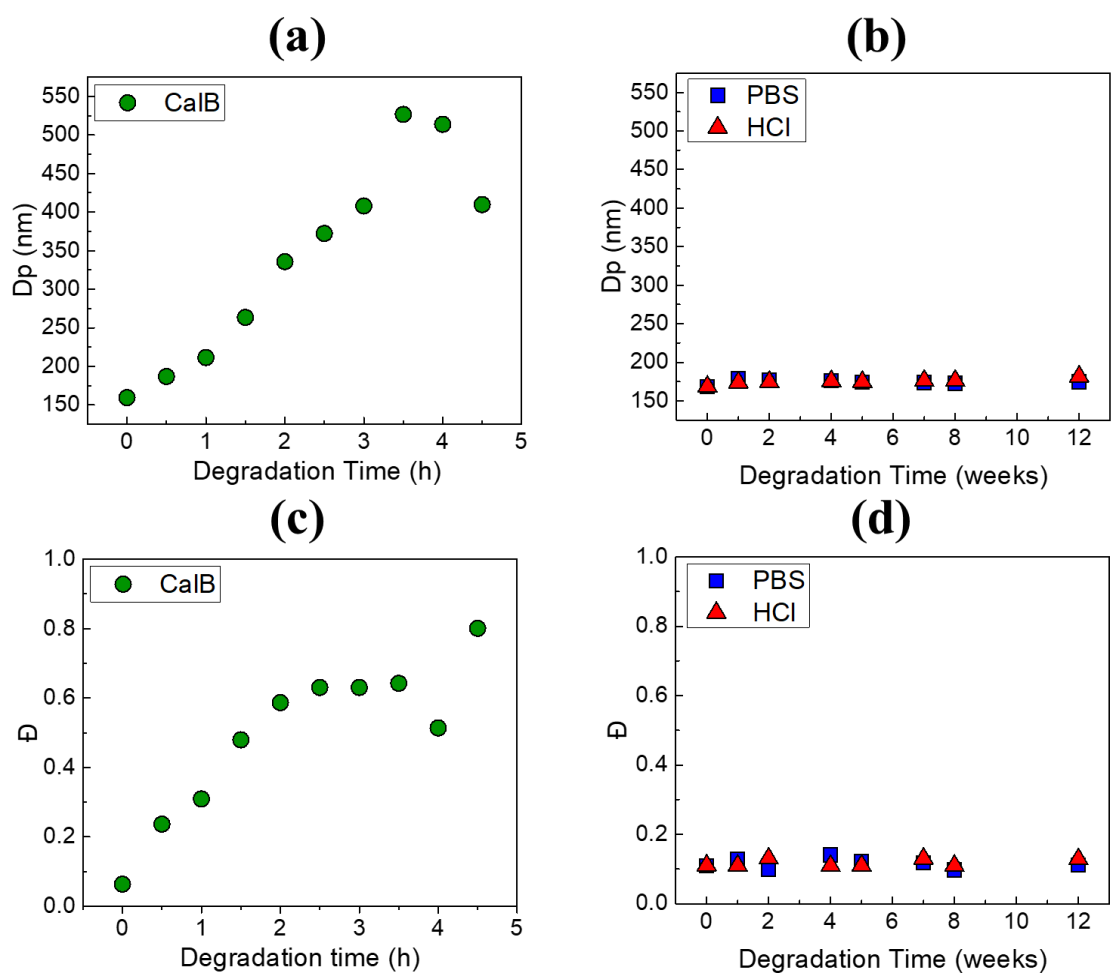

**Fig. 3** Particle size (a and b) and dispersity (c and d) of A-PTEe nanoparticles during degradation experiments in enzymatic solution (CalB, green circles), PBS solution (blue squares) and acidic solution (HCl, red triangles)

By ATR-FTIR analysis (Fig. 4), it was possible to identify the functional groups for the initial polymer and after degradation in different media. Before degradation in enzymatic solution (Fig. 4a), the sulfide bonds (-S-C-) can be confirmed at 720 cm$^{-1}$ demonstrating the successful thiyl radical addition to the double diene bond. A-PTEe samples were also analyzed by $^1$H NMR and the success of the reaction was confirmed. Fig. S3 (support information) presents A-PTEe spectra, that shows the complete consumption of the monomer double bonds, resulting in an yield higher than 99%, usual in step-growth polymerizations. The ether bonds stretching (-C-O-) are assigned to the band at 1098 cm$^-$ [1], while the stretching of (-C-O-) and (-C=O) bonds of the ester groups are observed respectively at 1170 cm$^{-1}$ and 1730 cm$^{-1}$ bands. After only 5 min incubation in enzymatic solution, it was possible to observe the decrease in the intensity of the band relative to the ester bond at 1770 cm$^{-1}$, which is an indicative of the hydrolysis of the ester bonds. At the same time, a new band appears at 1610 cm$^{-1}$, which can be assigned to the presence of deprotonated carboxylic acid, and a wide hydroxyl band is formed 3400 cm$^{-1}$. Fig. 4b shows ATR-FTIR spectra of A-PTEe nanoparticles before degradation and after the first and last week of degradation in PBS and HCl solutions. As expected, the band at 720 cm$^{-1}$ (-S-C-) remains unchanged after degradation in both PBS and HCl solutions. After the 12$^{th}$ week of incubation, it was possible to notice the occurrence of new bands at 1610 cm$^{-1}$ and 3400 cm$^{-1}$ and the decrease in 1170 cm$^{-1}$ and 1730 cm$^{-1}$ bands, in the same way as observed for degradation in enzymatic solution. However, for PBS and HCl solutions the changes in the spectra that indicate the degradation were much more subtle. These results, allied to the degradation curves of the samples and the theoretical results confirms that the degradation of A-PTEe nanoparticles occurs by the action of the water molecule, through hydrolysis reactions.

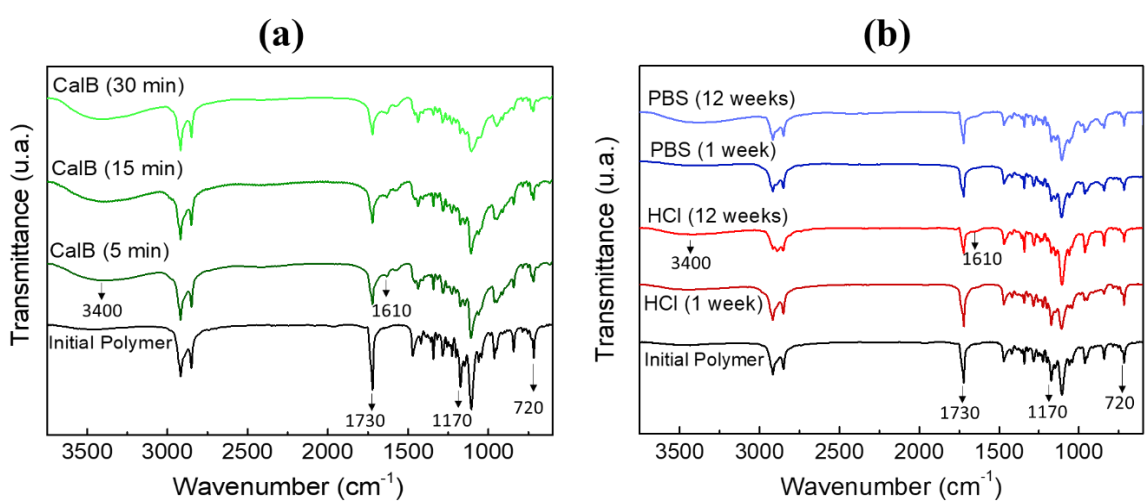

**Fig. 4** FTIR-ATR spectrum of A-PTEe (black line) and after degradation in (a) enzymatic solution (CalB, green lines) and (b) PBS solution (blue lines) and acidic solution (HCl, red lines)

Fig. 5 presents DSC thermographs for A-PTEe nanoparticles before and after degradation in the different tested media. A-PTEe presents two melting peaks, which means it have two distinct types of crystalline regions. The effect of the degradation in the crystalline regions of A-PTEe is evident, especially for samples degraded in enzymatic solution. In this case, after only 30 min of incubation, the crystalline arrangements were almost completely undone, remaining only a small peak at 36 °C, which persisted for the next 48 h of incubation, which might be composed of chain segments in which the ester bonds are not easily accessed by the enzymes and water molecules.

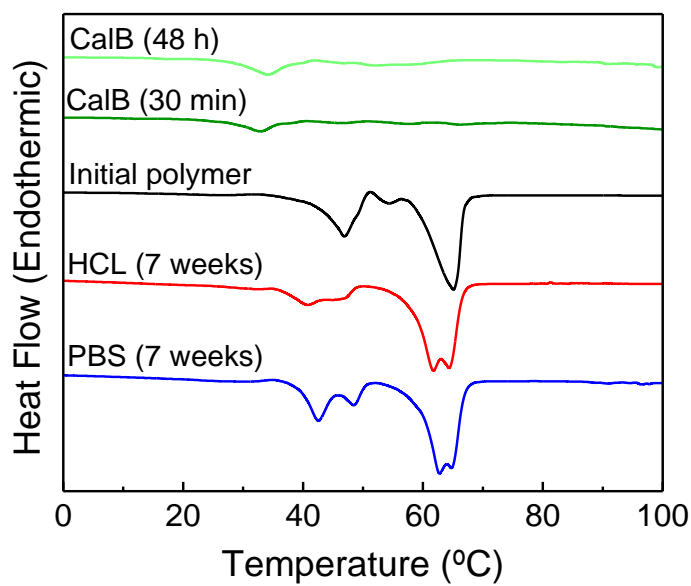

**Fig. 5** Thermograms of A-PTEe polymer before (black line) and after its degradation in PBS solution (7[th] week, blue line), acidic solution (HCl, 7[th] week, red line) and enzymatic solution (CalB, 30 min and 48 h, green lines)

When incubated in acidic solution, the effect of degradation in the crystalline region of A-PTEe at 50 °C was more evident in comparison to the A-PTEe nanoparticles degraded in PBS solution. As expected, the faster hydrolysis catalyzed by HCl results in more intense destruction of the crystalline domains, when comparing to degradation in PBS. In PBS solution however, a small peak could be seen at approximately 43°C indicating that degradation occurs more slowly, requiring more than 12 weeks for complete degradation of the polymeric material.

### 3.3. Citotoxicity assays

Due to its faster degradation behavior in comparison to B-PTEe nanoparticles, the cytotoxic effect of A-PTEe nanoparticles and their degradation products at different concentrations was investigated in L929 and human red blood cells. As shown in Fig. 6, A-PTEe nanoparticles did no present any cytotoxic effect on the L929 cells at concentrations of 10 to 100 µg/mL.

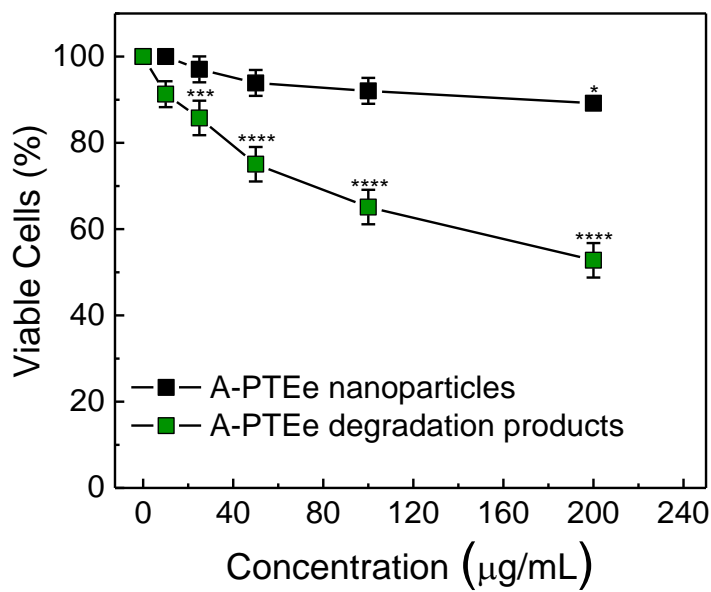

**Fig. 6** *In vitro* cytotoxicity of A-PTEe nanoparticles and its degradation products at different concentrations, in L929 murine fibroblast cells. Significant differences are shown (***p < 0.001, ****p < 0.0001 when compared with control group - Two-way ANOVA followed by Tukey test)

Only in the higher concentration PTEe nanoparticles presented a slight cytotoxicity on the L929 cells, decreasing the viability to 89%. On the other hand, their degradation products presented a significant cytotoxic effect at concentrations of 50, 100 to 200 µg/mL, decreasing the viability to 78, 72 and 58%, respectively. According to ISO 10993-5 [23] when the viability is ≥ 70 % the material is considered non-cytotoxic. Therefore, we can suggest that the concentration of 100 µg/mL of the degradation products generated by A-PTEe nanoparticles can be considered as non-cytotoxic.

The hemolysis assay is an important test for to evaluate the hemolytic potential of new materials for parenteral drug delivery [24]. The hemocompatibility of the product degradation of A-PTEe nanoparticles was evaluated on RBCs. As observed in Fig. 7, A-PTEe degradation products did not present any damage to the RBCs for all concentrations tested (up to 200 µg/mL).

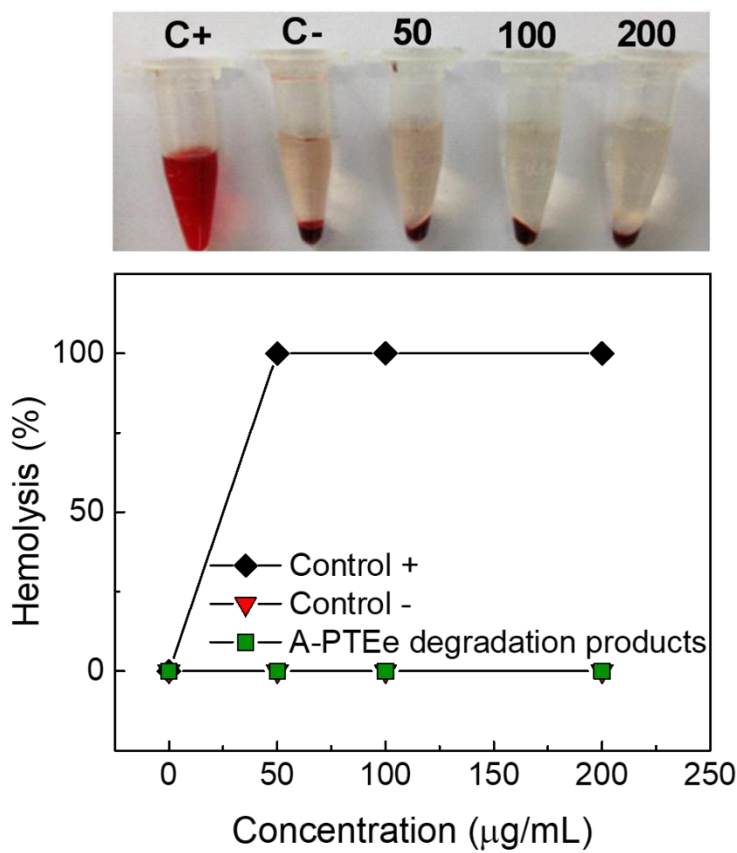

**Fig. 7** Hemolysis assay on RBCs upon incubation with the degradation products of A-PTEe nanoparticles at different concentrations. Data represents mean ± SD (n = 3)

Recent work has shown the hemocompatibility of A-PTEe nanoparticles [12]. These results reinforce the biocompatibility of the A-PTEe nanoparticles and its degradation products when exposed on fibroblast cells and RBCs.

## 4. CONCLUSION

A-PTEe nanoparticles presented rapid degradation behavior in aqueous media, especially when exposed to acidic and enzymatic solutions. In comparison to B-PTEe nanoparticles, molecular simulations revealed that A-PTEe has a stronger hydrophilic character (lower partition coefficient), explains its degradation behavior, determined experimentally. Biological assays also revealed that A-PTEe nanoparticles and its degradation products are biocompatible to murine fibroblast cells and RBCs. The data presented in this work highlight the potential of PTEe nanoparticles as potential materials for biomedical applications, and thiol-ene polymerization as a simple strategy that enables the green synthesis of these versatile materials from renewable sources.

**Acknowledgments**

We gratefully acknowledge CAPES (Coordenação de Aperfeiçoamento de Pessoal de Nível Superior), especially to CAPES-PRINT Program (Project number 88887.310560/2018-00), and CNPq (Conselho Nacional de Desenvolvimento Científico e Tecnológico) for the financial support. G. Candiotto and C. Guindani gratefully acknowledge FAPERJ (Fundação Carlos Chagas Filho de Amparo à Pesquisa do Estado do Rio de Janeiro), process number E-26/201.911/2020, for the financial support.

**Supporting Information**

A-PTEe and B-PTEe molecular structures optimized using DFT. $^1$H NMR spectrum for A-PTEe samples. Molecular weight distribution curves of A-PTEe nanoparticles after different degradation times in enzymatic solution. Molecular weight data for A-PTEe degraded in PBS after different degradation times.

**DECLARATIONS**

**Funding**

**Conflicts of interest/Competing interests**

The authors declare that they have no known competing financial interests or personal relationships that could have appeared to influence the work reported in this paper.

**Authors contributions**

F. Hoelscher: Investigation, Methodology, Writing - original draft, Writing - review & editing. P.B. Cardoso: Investigation, Concept development, Writing - original draft. G. Candiotto: Formal analysis, Investigation, Methodology, Writing - original draft, Writing - review & editing. C. Guindani: Formal analysis, Concept development, Methodology, Writing - original draft, Writing - review & editing. P. Feuser: Investigation - biological assays, Methodology, Writing - original draft. P.H.H. Araújo: Formal analysis, Conceptualization, Project administration, Supervision, Writing - review & editing. C. Sayer: Conceptualization, Writing - review & editing, Project administration, Funding acquisition, Supervision.

Nanomedicine Towards Cancer : J Pharm Sci 101:2271–2280.

https://doi.org/10.1002/jps